\documentclass[aps,prl,twocolumn,amssymb,showpacs,nofootinbib]{revtex4}
\usepackage{graphicx}
\usepackage{amsmath}
\usepackage{amssymb}
\usepackage{bm}

  \def\be{\begin{equation}}
  \def\beq{\begin{equation}}
\def\ee{\end{equation}}
\def\eeq{\end{equation}}
\def\bea{\begin{eqnarray}}
\def\eea{\end{eqnarray}}

\def\lam{\lambda}

\def\sq2{\sqrt{2}}

\def\ph{\varphi}

\def\m4{m^4(\ph)}
\def\mn2{m_n^2}

\def\v5{V^{(5)}}

\begin{document}

\title{\begin{flushright}
       \mbox{\normalsize \rm SU-4252-864}
       \end{flushright}
       \vskip 20pt
Metastable Kinks in the Orbifold}
\author{Manuel Toharia}
\author{Mark Trodden}

\affiliation{Department of Physics, Syracuse University, Syracuse, NY 13244, USA}
\date{\today}

\begin{abstract}
We consider static configurations of bulk scalar fields in extra dimensional models in which the fifth dimension is
an $S^1/Z_2$ orbifold. There may exist a finite number of such configurations,
with total number depending on the size of the orbifold interval. We perform a
detailed Sturm-Liouville stability analysis that demonstrates that all but the
lowest-lying configurations - those with no nodes in the interval - are
unstable. We also present
a powerful general criterion with which to determine which of these nodeless
solutions are stable. The detailed analysis underlying the results presented
in this letter, and applications to specific models, are presented in a
comprehensive companion paper~\cite{Toharia:2007xf}. 
\end{abstract}

\maketitle

The possibility of extra spatial
dimensions~\cite{Rubakov:1983bb,Akama:1982jy,Antoniadis:1990ew,Lykken:1996fj,Arkani-Hamed:1998rs,
Antoniadis:1998ig,Randall:1999ee,Randall:1999vf}, hidden from our current
experiments and observations through compactification or warping, has opened
up a wealth of options for particle physics model building and allowed
entirely new approaches for addressing cosmological problems.

In many models, standard model fields are supposed to be confined to a
submanifold, or brane, while in other models they populate the entire
bulk. Common to both approaches, however, is the inclusion of bulk fields
beyond pure gravity, either because they are demanded by a more complete
theory, such as string theory, or because they are necessary to stabilize the
extra dimensional manifold. The allowed configurations of such bulk fields are
determined, naturally, by their equations of motion, subject to the boundary
conditions imposed by the particular extra-dimensional model under
consideration. These might be periodic boundary conditions, in the case of a
smooth manifold, or reflection-symmetric ones in the case of an orbifolded
extra dimension.

In this letter we study a class of allowed nontrivial 
scalar field configurations~\cite{ArkaniHamed:1999dc,Georgi:2000wb,Kaplan:2001ga,Grzadkowski:2004mg} in
orbifolded extra-dimensional models, neglecting gravity. These
configurations exist whenever the potential possesses
at least two degenerate minima
and we show that they may form a finite tower of kink state solutions. We explicitly demonstrate
that all but the lowest-lying of these - the ones with no nodes in the
interval - are unstable. In addition we identify a stability criterion for these
lowest-lying states, and provide concrete examples for specific convenient
choices of potential.

A complete understanding of the predictions and allowed
phenomenology of extra dimension models necessarily includes a comprehensive
consideration of the configurations of bulk fields. That a finite tower of 
nontrivial static configurations may exist, with the possibility of
multiple stable ones, allows for new phenomena and
constraints on the models, and may have wide-ranging
implications for the particle physics and cosmological theories constructed
around them. 

We are currently performing a much more detailed analysis of these
configurations, including gravitational effects.

The model we consider consists of
a real scalar field in 5 dimensions (labeled by indices $M,N,\ldots =0,1,2,3,5$) around a flat background metric, and
defined by the action
\beq
S=\int d^5x\, \left[{1\over 2} \eta^{MN} \partial_M \phi(x,y) \;\partial_N \phi(x,y) - V (\phi)\right]
 \ ,
 \label{scal_models_general}
\eeq
in which $\phi$ has units of $({\rm Mass})^{3/2}$.
The extra dimension is compactified on an orbifold $S_1/Z_2$ with the scalar
field $\phi(x,y)$ being odd under $Z_2$ reflections along the extra coordinate
$x^5\!\equiv\!y$ (i.e. $\phi(x,y)=-\phi(x,-y)$). Here the orbifold
interval is defined as $[0,\pi R]$, with its size $\pi R$ assumed to be fixed.

The potential $V(\phi)$ must then be invariant under the discrete symmetry
$\phi\to-\phi$, and is chosen to have two degenerate global minima at
$\phi\!=\!\pm v$ with $v\!\neq\!0$. To simplify notation, we will also choose the potential 
to vanish at $\phi\!=\!0$.

We seek static field configurations $\phi_{{}_A}(y)$, parametrized by their amplitudes $A$,
which extremize the action, and with nontrivial $y$-dependence, subject to the appropriate boundary conditions,
namely $\phi_{{}_A}(0)\!=0\!$ and $\phi_{{}_A}(\pi R)\!=\!0$. 
The field equation satisfied by such solutions is
\bea
\phi_{{}_A}{\!\!\!''}- {\partial V \over \partial \phi_{{}_A}} = 0 \ ,
\label{scal_el_general}
\eea
where a prime denotes a derivative with respect to $y$.
It is easily seen that there exists a first integral, given by 
\beq
\frac12 \phi_{{}_A}{\!\!'}^2-V(\phi_{{}_A})=E_A \ .
\label{energy}
\eeq

\begin{figure}[h]
\begin{center}
\includegraphics[width = 8cm]{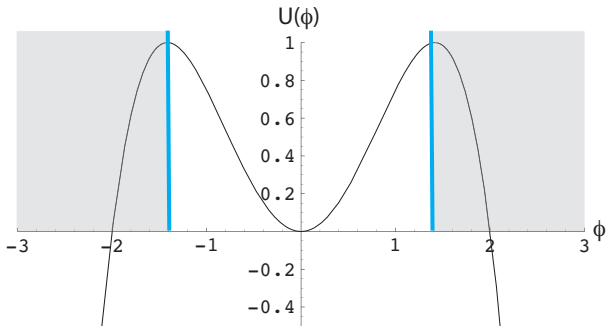}
\end{center}
\vspace{-.5cm}
\caption{Mechanical Analogy: Periodic solutions of a particle in the potential
  $U(\phi)=-V(\phi)=(\mu^2/2)\phi^2-(\lambda/4)\phi^4$ (here with $\mu^2=2$ and $\lambda=1$) exist when the
  total energy of the particle lies between $E_{\rm max}={\mu^4\over 4\lambda}$ (top of the inverted potential) and
  $E_{min}=0$. A particle with energy $E_{{}_A}$ will undergo a
  periodic motion of period $P$, understood as the length of the extra-dimension. Note that this is precisely the potential 
  used in our first example~(\ref{v5}).}
\label{mechanicalanalog}
\vspace{.0cm}
\end{figure}

To better understand the problem, it is extremely helpful to
note~(Figure~\ref{mechanicalanalog}) that~(\ref{scal_el_general})
and~(\ref{energy}) are precisely the equations of a particle rolling in time ($y$),
without friction, in the inverted potential $U(\phi)\equiv-V(\phi)$.

By inspection of Figure~\ref{mechanicalanalog}, whenever the
inverted potential $U(\phi)$ has a global maximum not at $\phi=0$, there
will always exist a range of values for the constant of integration $E_{{}_A}$ such
that at least one nontrivial solution, with appropriate boundary
conditions, exists. Each solution $\phi_{{}_A}(y)$ will actually be periodic, with period $T(A)$, and
will reach its maximum value $A$ at $y=T/4$. The constant parameter $E_{{}_A}$ of
each solution is set by the amplitude $A$ of the solution since
$E_{{}_A}\!=\!-V(A)$. Using~(\ref{energy}), we may also derive an expression for the period
$T$ as a function of $A$
\be
T(A) = 2\sqrt{2}\int_0^A \frac{d\phi}{\sqrt{V(\phi)-V(A)}} \ .\label{TA}
\ee

It is straightforward to show that there is a family of possible nontrivial periodic solutions
parametrized by the amplitude $A$.
To relate the period $T(A)$ of a solution to the physical size of
the interval $\pi R$ one simply notes\cite{Grzadkowski:2004mg} that one has to
be be a multiple of the other, i.e. $\ 2\pi R=(m+1)\ T(A)$, where $m$ is
a positive integer. In that case, the nontrivial solution will contain $m$
nodes inside the interval $[0,\pi R]$.
As long as the constant $E_{{}_A}$ has a value in the appropriate range, there
will always be at least one nontrivial solution with $m\!=\!0$, i.e. nodeless
in the orbifold interval. Depending on the function $T(A)$, there may be more
than one value of $A$ such that $\ 2\pi R= T(A)$, and therefore more than one
nontrivial nodeless solution.

On the other hand the function $T(A)$ must have a non-zero global
minimum $T_{\rm min}$.
Once $T_{\rm min}$ (determined by the specific potential
$V(\phi)$ chosen) as well as the size of the orbifold interval $\pi R$ are known, then
defining $2\pi R/T_{\rm min}=\delta$, the maximum number of nodes that a nontrivial solution $\phi_{{}_A}(y)$ can have
is $m_{{}_{max}}=IP(\delta)-1$ (where $IP(x)\equiv$ IntegerPart$(x)$ gives
the largest integer less than or equal to $x$.), and $m_{{}_{max}}\!+1$ is the
maximum number of solution periods that can ``fit'' in $2 \pi R$.

This means that a finite number of nontrivial solutions with nodes 
are also possible.

The most important question about these background solutions concerns stability. To investigate this,
we add small perturbations around one such solution $\phi_{{}_A}(y)$ of~(\ref{scal_el_general}), i.e. 
we write $\ \phi(x,y)\to \phi_{{}_A}(y) +\varphi(x,y)$. Separation of variables $\varphi(x,y)=\varphi_x(x)\varphi_y(y)$ then yields
\bea
&&\Box \varphi^n_x(x)\ =\  -M_n^2\ \varphi^n_x(x)\hspace{1cm}\\
&&{\varphi_y^n}^{''}(y) - q(y)\varphi^n_y(y)\ =\ -M_n^2\ \varphi^n_y(y) \ ,
\label{kkmodes}
\eea
with
\be
q(y) = {\partial^2 V\over \partial
    \phi^2} \Big|_{\phi_{{}_A}(y)} \ ,
\ee
which are the equations of motion of a (Kaluza-Klein) tower of 4D scalar fields $\varphi^n_x(x)$
with masses $M_n^2$ and with profiles along the extra dimension given by
$\varphi^n_y(y)$.

The masses $M_n^2$ and profiles $\varphi^n_y(y)$ are found by solving the second
order linear differential equation (\ref{kkmodes}).

Now, suppose we have identified a solution $\phi_{{}_A}(y)$ of equation~(\ref{scal_el_general}). Taking the 
derivative of~(\ref{scal_el_general}), one obtains
\bea
\phi_{{}_A}'''(y) -q(y) \phi_{{}_A}'(y)=0 \ .
\label{kinkder}
\eea

Comparing Equations (\ref{kinkder}) and (\ref{kkmodes}) we then see that the
derivative $\phi_{{}_A}'(y)$ of the background solution actually corresponds to a
massless mode ($M_n^2=0$) of the perturbation $\varphi(x,y)$, although now with Neumann boundary
conditions, rather than the Dirichlet ones we require\footnote{This should not come as
a big surprise, since it is just the translation mode, the masslessness of
which is a reflection of translation symmetry. It cannot be a physical solution
since translation invariance is broken in the orbifold.}.

This is extremely useful, however, since the general theory of the eigenvalues
of the Sturm-Liouville problem with Dirichlet (D), 
periodic (P), semiperiodic (S), and Neumann (N) boundary conditions implies the following chain of inequalities
\bea
\lambda_0^N &\leq& \lambda_0^P <\lambda_0^S \leq \{\lambda_0^D,\ \lambda_1^N\}\leq \lambda_1^S <\lambda_1^P
\leq \{\lambda_1^D,\ \lambda_2^N\} \nonumber \\
&\leq& \lambda_2^P <\lambda_2^S \leq \{\lambda_2^D,\ \lambda_3^N\}\leq \lambda_3^S <\lambda_3^P
\leq \{\lambda_3^D,\ \lambda_4^N\} \nonumber \\
&\leq& \cdots \ ,
\label{eigenvalueinqual}
\eea
relating the towers of eigenvalues corresponding to each different type of boundary condition.

Applying this to any solution $\phi_{{}_A}(y)$ with greater than the minimal periodicity, we see that the associated derivative 
$\phi_{{}_A}'(y)$, obeying Neumann boundary conditions, will have multiple nodes 
in the interval $[0,\pi R]$. Thus we may identify it as the eigensolution $\varphi_i^N(y)$, with $i\geq 2$, with its masslessness 
implying that the corresponding eigenvalue obeys $\lambda_i^N =0$. 

However~(\ref{eigenvalueinqual}) implies that $\lambda_2^N>\lambda_0^D$. Therefore, if $\lambda_i^N=0$ for some $i\geq 2$,
then there exists at least one ($\lambda_0^D$) eigenvalue of the Dirichlet problem, and possibly more, that are negative!

Thus, {\it all static solutions with nodes in the interval are unstable}.

What remain are the lowest-lying solutions, with no nodes in 
$(0,\pi R)$. Note again that, depending on the complexity of the potential, there
may be multiple static, nodeless solutions with the same periodicity. A different approach is 
necessary to investigate their stability, and here we shall merely state the
result. It turns out that the key ingredient is the dependence of the
periodicity of static solutions T on their amplitude $A$.

It is possible to prove the following general result. {\it A static,
  nodeless solution $\phi_{{}_{A_*}}(y)$ to equation~(\ref{scal_el_general}),
with amplitude $A_*$, and period $T(A_*)$, and satisfying
  $\phi_{{}_{A_*}}(0)=\phi_{{}_{A_*}}(T/2)=0$, is stable if and only if
\beq
\left.\frac{dT}{dA}\right|_{A=A_*} > 0 \ .
\label{stabilitycriterion}
\eeq
} This is the central result presented in this letter, and we reserve the
somewhat involved proof for our companion paper~\cite{Toharia:2007xf}.

Further,  since the energy of such a solution is given by
\be
E(A)=V(A)T(A)+4\sqrt{2}\int_0^A \sqrt{V(\phi)-V(A)}\, d\phi \ ,
\label{energyvsA}
\ee
when there are multiple metastable nodeless solutions with the same value of $T$, it is straightforward to compare their
energies and determine which is the lowest lying state.

After all this generality, it may prove useful to sketch how some of this works out in some concrete examples.

Let us first consider an exactly solvable example defined by the potential
\beq
V(\phi)= -{\mu^2 \over 2} \phi^2 +{\lambda \over 4} \phi^4\ ,
\label{v5}
\eeq
where $[\mu]=[\lambda]^{-1}=({\rm Mass})$.

It is easy to see that for $E={\mu^4\over 4\lambda}$ one obtains non-trivial solutions of Eq.~(\ref{energy}) known 
as the kink and anti-kink
\beq
\phi_{\rm (anti-)kink}(y)= \pm {\mu \over \sqrt{\lam}} \tanh\left[{\mu \over \sqrt{2}}\ (y-y_o)\right] \ ,
\label{kink}
\eeq
where the kink location $y_o$ should be set to zero because of the boundary
conditions of the scalar field.
This solution interpolates along the (now infinite) extra dimension between the
constant background solutions $\phi_{\pm}\equiv \pm{\mu/ \sqrt{\lam}}$.

For $0<E<{\mu^4\over 4\lambda}$, we can still integrate Eq.~(\ref{energy})
to obtain~\cite{Grzadkowski:2004mg}
\beq
\phi_k(y)=\pm 
{\mu\over \sqrt{\lambda}} \sqrt{2 k^2 \over k^2+1}\   {\rm sn}\left({\mu\over
    \sqrt{k^2 +1}}\ y ,\ k^2 \right) \ ,
\label{periodicvev}
\eeq
where 
\beq
k^2
={\mu^2-\sqrt{\mu^4-4 \lam E} \over \mu^2+\sqrt{\mu^4-4 \lam E} }
\eeq
and ${\rm sn}(x,k^2)$ is the Jacobi Elliptic Sine-Amplitude, parametrized by the elliptic modulus $k$
(a real parameter such that $0<k<1$) and with period $4 K$, where
\bea
K(k^2)=\int_0^{\pi/2} {d\theta \over \sqrt{1 - k^2 \sin^2\theta}}
\eea
is the complete elliptic integral of the first kind.

In this example, it is possible to show that the total number $n_{\rm
  max}$ of nontrivial solutions is given by 
$n_{\rm max}={\rm IP}(\mu R) -1$.
Since $\mu$ is a fixed parameter of the scalar potential and $R$ is the fixed radius of the
extra dimension, $n_{\rm max}$ is completely specified by the model.

The complete set of static nontrivial background solutions consistent
with the boundary conditions, for the potential~(\ref{v5}) is then 
\be
\phi_{k_n}(y)=\pm {\mu\over \sqrt{\lambda}} \sqrt{2 k_n^2 \over k_n^2+1}\
  {\rm sn}\left({\mu \over \sqrt{k_n^2 +1}}\ y ,\ k_n^2 \right) \ ,
\ee
where $n$ is an integer such that $\ 0 \le n \le n_{\rm max}$. 

The radius $R$ of the extra dimension is such that $2\pi R={4\over \mu}\sqrt{k_0+1}\ K(k_0^2)$.

\begin{figure}[h]
\begin{center}
\includegraphics[width = 8.5cm]{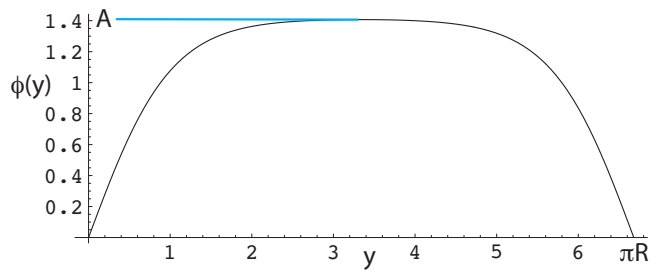}
\end{center}
\vspace{-.5cm}
\caption{The single stable kink solution for the case of the potential of our first example~(\ref{v5}).}
\label{phi4kink}
\vspace{.0cm}
\end{figure}

The solution with lowest
energy, and no nodes in the interval, will be $\phi_{k_0}(y)$ (using $k$ as an equivalent label to $A$)
and is plotted in Fig.~(\ref{phi4kink}). The rest of 
solutions $\phi_{k_n}(y)$ will 
have nodes and increasing energy. And thanks to our general stability argument
they will be unstable.

A second example, useful for visualizing the possibility of multiple stable solutions, is provided by the (admittedly contrived)
potential
\be
V(\phi)=-\phi^2 -5\phi^4+\frac{5}{2}\phi^6 -\frac{1}{3}\phi^8+\frac{1}{77}\phi^{10} \ ,
\label{secondexample}
\ee
in which we have set all dimensionful parameters to unity.

\begin{figure}[h]
\begin{center}
\includegraphics[width = 8.5cm]{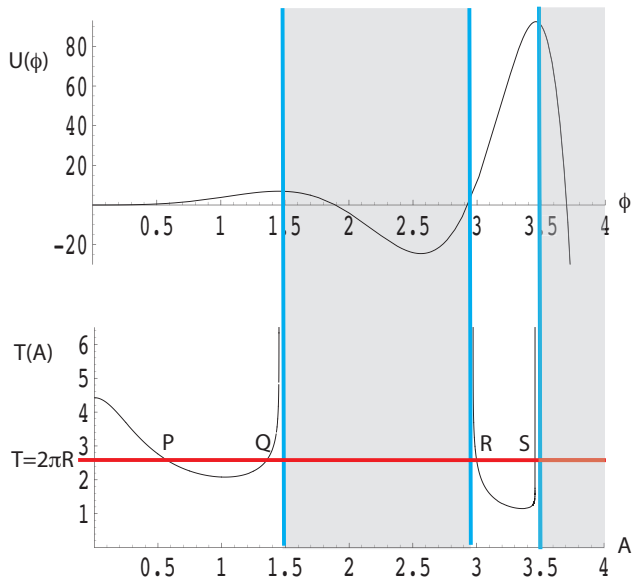}
\end{center}
\vspace{-.5cm}
\caption{For this example of the potential, there exist four distinct nodeless solutions, here labeled as $P$, $Q$, $R$ and $S$, 
with  different values of the amplitude $A$, but with the same period. Those at $P$ and $R$ are unstable, while those at $Q$ 
and $S$ are stable. Further,~(\ref{energyvsA}) implies that $S$ is of lower energy than $Q$. In the shaded regions there are 
no solutions with the appropriate boundary conditions.}
\label{stability}
\vspace{.2cm}
\end{figure}

Obviously, our general considerations imply that the only possible stable configurations are the nodeless ones. 
In figure~\ref{stability} we plot the inverted potential $U(\phi)$ for $\phi>0$ and the period $T(A)$ of static solutions
with amplitude $A$. 

\begin{figure}[h]
\begin{center}
\includegraphics[width = 8.5cm]{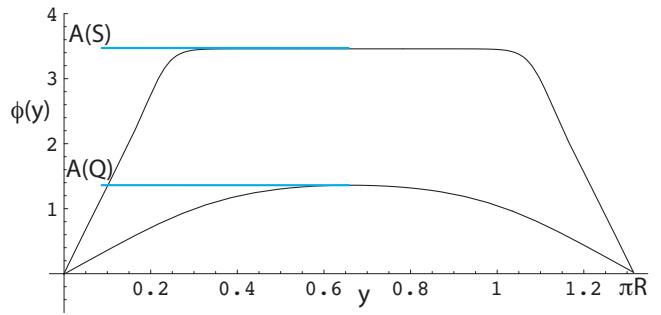}
\end{center}
\vspace{-.5cm}
\caption{The two stable solutions (points $Q$ and $S$ in
  figure~(\ref{stability})) for the potential~(\ref{secondexample}). The
  solution at point $S$, with the larger amplitude, $A(S)$, has the lower energy.}
\label{QSsolutions}
\vspace{.2cm}
\end{figure}

If we are looking for nodeless solutions,
we are interested in those values of $A$ satisfying $T(A)=2\pi R$ and by
inspection there are four such solutions, labeled $P$, $Q$, $R$ and $S$. Furthermore, our
stability criterion~(\ref{stabilitycriterion}) then implies that $P$ and $R$
are unstable, whereas $Q$ and $S$ are distinct, stable, nodeless
solutions. These two solutions are plotted in Fig.~(\ref{QSsolutions})

In addition, using~(\ref{energyvsA}), one can show that the solution denoted by $S$ is the lowest energy one.

In this letter we have investigated the existence of a finite tower of
nontrivial static configurations in orbifolded extra dimension models
containing bulk scalar fields with potentials with multiple degenerate
vacua. We have described a proof that all but the lowest-lying of these
configurations are classically unstable, and have discussed a general
stability criterion for the lowest, nodeless, modes. We have then illustrated
these general results first with the case of an exactly solvable model and
then with a more complicated but also richer example.

We have thus analyzed an entirely new class of metastable bulk field
configurations around which novel particle physics and cosmological phenomena
may occur. The stability criterion~(\ref{stabilitycriterion}) of the lowest lying modes requires an
involved proof, which we reserve for our companion paper, in which we also
present a detailed discussion of the configurations described in this
letter. We also apply these considerations to a range of models. 

Natural further steps are the inclusion of gravity into the analysis, and a
comprehensive investigation of the applications of our results. These studies are already underway.

We thank Tim Tait and James Wells for discussions.
M.Toharia is supported by funds provided by Syracuse University and the U.S. DOE 
under Contract number. DE-FG-02-85ER 40231. 
M. Trodden is supported by the National Science Foundation under 
grant PHY-0354990 and by Research Corporation.

\end{document}